\documentstyle{article}

\newcommand{\g}{\varepsilon_g}
\newcommand{\e}{\delta c^2}
\date{}
\begin{document}

\chardef\active=13 \catcode`\"=\active
\def"#1{\if a#1\accent127a%
\else	\if u#1\accent23u%
\else	\if o#1\^o%
\else	\if A#1\accent127A%
\else	\if U#1\accent23U%
\else	\if O#1\^O%
\else	\if d#1{d\kern-.035cm'\kern-.03cm}%
\else	\if l#1{l\kern-.035cm'\kern-.03cm}%
\else	\if t#1{t\kern-.035cm'\kern-.03cm}%
\else	\if L#1{L\kern-.08cm'}%
\else\accent20#1\fi\fi\fi\fi\fi\fi\fi\fi\fi\fi}

\chardef\active=13 \catcode`\`=\active
\def`#1{\ifmmode{\cal#1}%
\else	\if i#1\'{\i}%
\else	\if '#1{\strut\raise0.3ex\hbox{\char34}}%
\else	\if\string`\string#1\char92%
\else	\if,#1{\strut\lower1.2ex\hbox{\char34}}%
\else\'#1\fi\fi\fi\fi\fi}

\title{Gravitation Interaction and Electromagnetic Interaction in the 
Relativistic Universe with Total Zero and Local Non-Zero Energy\thanks
{This work was supported by Intertextil Praha.}}

\author{ Vladim`ir Skalsk`y$^1$    
and  
Miroslav S`uken`ik$^1$}
\maketitle

\begin{abstract} 
In the model of flat expansive homogeneous and isotropic relativistic 
universe with total zero and local non-zero energy the gravitation energy 
of bodies and the elecromagnetic energy of charged bodies can be localised.
\end{abstract}

\vspace{.5cm}

\section{INTRODUCTION}

The equations of the Einstein gravitation law [1] can be expressed in the 
form:

$$ R_{ik} - \frac12 g_{ik}R = \frac{8\pi G}{c^4}T_{ik}, \eqno(1) $$
where $ R_{ik} $ is the curvature tensor, $g_{ik} $ is the metric tensor,
$ R$ is the scalar curvature, and $ T_{ik} $ is the energy-momentum tensor.

If we can determine the gravitation energy density in strong fields we 
must use the Isaacson tensor; in weak fields we get along with the Tolman
approximation.

From equations (1) results relation for the gravitation energy density of 
matter objects $\g$ in the first (Tolman's) approximation:

$$ \g = - \frac{c^4}{8\pi G}R. \eqno(2) $$

From observations results that $\g$ is a non-zero quantity.

The first exact solution of equations (1) was found Schwarzschild in 1915. 
In the Schwarzschild metrics [2] the scalar curvature $R$ outside of the 
area of the matter object has the value:

$$ R = 0. \eqno(3)$$

From relations (2) and (3) it results:

$$ \g = 0. \eqno(4) $$
This fact was interpreted as a non-localisation of the gravitation energy.

\section{MODEL OF THE EXPANSIVE NONDECELERATIVE UNIVERSE}

The mathematical and physical fundamentals of {\it the standard model of 
universe} is represented by the Friedmann equations of the dynamics of 
homogeneous and isotropic relativistic universe [3, 4]. Using the 
Robertson-Walker metrics [5-9] they can be written in the form:

$$ \dot{a}^2 =\frac{8\pi G\rho a^2}{3} - kc^2 + \frac{\Lambda a^2c^2}{3}, \eqno(5a)$$

$$ 2a\ddot{a} + \dot{a}^2 = -\frac{8\pi Gpa^2}{c^2} - kc^2 + \Lambda a^2c^2, \eqno(5b)$$
where  $a$ is the gauge factor, $\rho $ is the mass density, $k$ is the 
curvature coefficient, $\Lambda $ is the cosmological member, and $ p $ is 
the pressure.

The Friedmann-Robertson-Walker equations (5a) and (5b) are the applications 
of the Einstein equations (1) to the whole of the universe in the Newton 
approximation, under 
supplementary assumptions introduced into relativistic cosmology by Einstein 
[10] and de Sitter [11] and generalized by Friedmann [3, 4].

The total energy of universe is exactly zero [12, p. 129]. In the General 
Theory of Relativity the curvature of space-time is determined by the energy 
density $\varepsilon$ plus three-multiple of the pressure $ p$. Therefore, in 
the (flat) expansive nondecelerative (homogeneous and isotropic relativistic)
 universe (with total zero and local non-zero energy) (ENU) holds the 
relation: $\varepsilon + 3p = 0,$ i.e. such a universe is determined by the 
state equation [13]:

$$ p = - \frac13 \varepsilon . \eqno(6)$$

The Friedmann-Robertson-Walker equations (5a) and (5b) describe all 
theoretically possible models of the homogeneous and isotropic relativistic 
universe, therefore, the model of ENU determined by the state equation (6) 
must be their solution.

The state equation (6) is the solution of equations (5a) and (5b) only with  
$ k = 0 $ and $\Lambda = 0.$ Therefore, the ENU is the only model of 
homogeneous and isotropic relativistic universe with total zero and local 
non-zero energy [13].

In the ENU - determined by the Friedmann-Robertson-Walker equations (5a) and 
(5b) with $ k = 0 $ and $\Lambda = 0 $ and the state equation (6) - hold 
relations [13-15]:

$$ a = ct = \frac{2GM}{c^2} = \sqrt{\frac{3c^2}{8\pi G\rho}} , \eqno(7)$$
where $ t $ is the cosmological time and $ M $ is the mass of ENU.

The relation (7) gives the relation for the (critical) energy density 
(of the whole ENU) [13]

$$ \varepsilon = \frac{3c^4}{8\pi Ga^2}. \eqno(8) $$

From the relations (7) results the increase of energy [16]

$$ \e = \frac{dM}{dt}c^2 = \frac{c^5}{2G} = 1.815 \times 10^{52} J s^{-1}. \eqno(9) $$

From the relations (7)-(9) follows that in the ENU the energy increase 
$\e $ (9) - which we had interpreted as the permanent constant maximum 
possible creation of energy [16] - is one from reasons that the (critical) 
energy density of the whole ENU $\varepsilon $ (8) decreases 
proportionally to $ a^{-2} $ (and not proportionally to $a^{-3}$). This is in
good agreement with observations [17].

For the mass of a homogeneous sphere $ m_x $ with the mass density 
$\rho $, the radius $ r_x $ and the volume $ V_x $ are valid the relations:

$$ m_x = \frac43\pi r^3_x\rho = V_x\rho . \eqno(10) $$

The escape velocity $ v_x $ from the surface of the homogeneous sphere with 
the mass $ m_x $ and the radius $r_x $ is:

$$  v_x = \sqrt{\frac{2Gm_x}{r_x}}. \eqno(11) $$

According to the relations (7), the ENU in the distance of the gauge factor 
$ a $ expands at the boundary velocity of signal propagation $ c$. Therefore, 
the velocity of radial expansion $ v_y $, corresponding to the distance $r_x$,
is given by the relation [17]:

$$ v_y = \frac{r_x}{a}c. \eqno(12) $$

From the relations (10)-(12) it follows [17]:

$$ \frac{2GV_x\rho a^2}{c^2} = r^3_x. \eqno(13) $$

From the relation (13) follows the relation (8).

The gravitation interaction theoretically acts in the range of the whole 
universe. However, from the relations (8) and (13) results that in the ENU in 
an arbitrary large sphere with the (critical) energy density of the whole ENU 
$\varepsilon $ (8) the gravitation interaction is exactly compensated by its 
expansion.

The gravitation energy density $\g $  in the Newton approximation can be 
expressed by the relation:

$$ \g = -\frac{({\it Grad} \varphi )^2}{8\pi G}. \eqno(14)$$

From the relations (8) and (14) it follows that in the ENU the gravitation 
action of a body can manifest itself only up to the distance in which the 
absolute value of gravitation energy density of its field $| \g | $ 
(14) is larger than the (critical) energy density of the whole ENU 
$\varepsilon $ (8) [17].

This means that in the ENU the equations (8) and (14) determine the effective 
gravitation range $ r_{ef} $ of an arbitrary body with the gravitation radius 
$ r_g $:

$$ r_{ef} \approx \sqrt{r_ga}. \eqno(15) $$

\section{LOCALIZATION OF THE GRAVITATION ENERGY AND THE GRAVITATION 
INTERACTION OF TWO BODIES IN THE ENU}

The relation (14) is a good approximation for the whole ENU, but it does not 
allow to localize the gravitation energy.

According to the relations (7)-(9), the energy in ENU is the function of time,
 therefore, its determination requires a metrics taking into account also the 
change of energy in time. Such a property has the metrics of Vaidya [18], 
which contains the metrics of Schwarzschild [2] as a special partial solution.

After introducing the Vaidya spinor components to the Riemann tensor and the 
Vaidya spinor coefficients to the Ricci tensor, the field metrics can be 
expressed using the quasi-spherical coordinates [19, p. 105]:

$$   ds^2 =  \left( 1 - \frac{2\Psi ^0_{(t)}}{r} \right) c^2dt^2 -  
\left( 1 - \frac{2\Psi ^0_{(t)}}{r}\right)^{-1}dr^2 - r^2\left( d\Theta^2 + 
\sin^2\Theta d\varphi^2\right) ,  \eqno(16) $$ 
where

$$ \Psi ^0_{(t)} = \frac{Gm_{(t)}}{c^2} , \eqno(17) $$
where $ m_{(t)} $ is a variation of mass with time.

In the Vaidya metrics [18] the scalar curvature R satisfies the relation [20]:

$$ R = \frac{6G\dot{m}_{(t)}}{r^2c^3} , \eqno(18) $$
where $ r $ is the distance in which we measure the scalar curvature $ R $.

Using the relations (7) and (9), we can express the relation (18) in the 
form [20]:

$$ R = \frac{3r_g}{r^2a} . \eqno(19) $$
 
Further, using the relation (19), we can express the relation (2) in the 
form [20]:

$$ \g = - \frac{3mc^2}{4\pi r^2a} . \eqno(20) $$

For the whole ENU (under conditions: $ m = M $ and $ r = a = r_g $) is valid 
the relation [20]:

$$ \g{}_{(ENU)} = - \frac{3c^4}{8\pi Ga^2}, \eqno(21) $$
where $ \g{}_{(ENU)} $ is the gravitation energy density of the whole ENU.

This means that the absolute value of gravitation energy density of the whole 
ENU $ | \g{}_{(ENU)} | $ (21) is exactly equal to the (critical) energy 
density of the whole ENU $ \varepsilon $ (8), i.e. the total energy of ENU 
$ E_{tot} $ is exactly equal to zero [21].

Comparing the relations (20) and (21) in the ENU, we receive the effective 
gravitation range $ r_{ef} $ of an arbitrary body with gravitation radius 
$ r_g $ [20]:

$$ r_{ef} = \sqrt{r_ga} .\eqno(22) $$
(Compare the relations (15) and (22).)

Let us now analyze the gravitation interaction of two bodies.

Let us have two bodies. The big one with mass $ m_1 $ and the small body with 
mass $ m_2 $ at a distance $ r $.

In the ENU  in the first approximation for the effective gravitation surface of 
the body with mass $ m_2 $ $ S_{ef(m_2)} $ is valid [20]:

$$ S_{ef(m_2)} = \pi r^2_{ef(m_2)} = \pi r_{g(m_2)}a . \eqno(23) $$

In the ENU in the first approximation the double multiple of the flow of 
gravitation energy density from the body with mass $ m_1 $ across the effective 
gravitation surface of body with mass $ m_2 $  $ S_{ef(m_2)} $ determines the 
gravitation force $ F_g $ acting between these bodies. (See Figure 1.)

The relations (2), (20) and (23) give [20]:

$$ F_g = - \frac{c^4}{4\pi G} R_{(m_1)} \frac{\pi r^2_{ef(m_2)}}3 = 
-\frac{Gm_1m_2}{r^2}, \eqno(24) $$
where $ R_{(m_1)} $  is the scalar curvature of gravitation field of the body 
with mass $ m_1 $ in the distance $ r. $

To the gravitation force $ F_g $ (24) corresponds the gravitation potential 
energy [20]

$$ E_g = - \frac{Gm_1m_2}{r}. \eqno(25) $$

\section{PROBLEM OF DARK MATTER IN THE ENU}

In the ENU the equation (22) expresses the relation between the mass and the 
dimensions of the gravitationally bound system.

This relation can be used for all hierarchical gravitationally bound 
rotational systems (HGRS) (planetary systems, clusters of stars, galaxies, 
clusters of galaxies, superclusters, and others) in the ENU.

If we know the gauge factor (of ENU) $ a$, the effective dimensions of a HGRS 
$r_{ef}$ 
and its visible mass $m_{obs}$ in the ENU, we can exactly determine the percentage 
of the dark matter in HGRS without measuring its rotational velocity. 
In the ENU hold the relations [20]:

$$ r^2_{ef} = r_ga = \frac{2Gm_{glob}a}{c^2}, \eqno(26) $$
where $ m_{glob} $ is the global mass of the HGRS.

From the relations (26) it results [20]:

$$ \frac{m_{obs}}{m_{glob}}100 = \frac{2Gm_{obs}a}{r^2_{ef}c^2}100 = 
\% \ \  of \ \  visible \ \ matter. \eqno(27) $$

From the relations (26) and (27) it results [20]:

$$  \left(1 - \frac{2Gm_{obs}a}{r^2_{ef}c^2}\right)100 = \% \ \  
of \ \  dark \ \  matter. \eqno(28) $$

{\it For example:} According to observations, the average supercluster has 
the diameter $ 2r_{ef} \sim 1.3 \times 10^8 $ ly and the visible mass 
$ m_{obs(sup)} \sim 10^{46} $ kg. From the relation (28) and the present 
gauge factor of Universe (our ENU) [22]

$$ a_{pres} = 1.299 \times 10^{26} {\it m} , \eqno(29) $$
it follows that there is approximately 99\% of dark matter in the average 
supercluster [20].

\section{ESSENTIAL UNITY OF GRAVITATION \newline
AND ELECTROMAGNETIC INTERAC- \newline TIONS IN THE ENU}

According to Shipov [19], the equations of field of geometrized 
electrodynamics can be expressed in the form [19, p. 112]:

$$ R_{ik} - \frac12 g_{ik}R = \frac{8\pi e}{mc^4}T_{ik} , \eqno(30) $$
where the tensor of energy-momentum $ T_{ik} $ can be expressed by using the 
density of electromagnetic energy $ \rho _ec^2 $:

$$ T_{ik} = \rho _ec^2u_iu_k . \eqno(31) $$

The metrics of field of the charge body can be written in the form 
[19, p. 113]:

$$ ds^2 = \left( 1 - \frac{r_e}r\right)c^2dt^2 - \left( 1 - 
\frac{r_e}r\right)^{-1}dr^2 - r^2 \left(d\Theta ^2 + \sin ^2\Theta 
d\varphi ^2\right), \eqno(32) $$
where the electromagnetic radius

$$ r_e = \frac{2e^2}{mc^2}. \eqno(33) $$

The energy density of the electromagnetic field $\varepsilon _e $ can be 
expressed by the relation [19, p. 116]:

$$ \varepsilon _e = -\frac{mc^4}{8\pi e}R , \eqno(34) $$
where $ R $ is the scalar curvature.

The monopol radiation of a charge can be described by the Vaidya metrics 
[19, p. 166]:

$$ ds^2 = \left( 1- \frac em \frac{2e_{(t)}}{rc^2}\right)c^2dt^2 - 
\left( 1+ \frac em \frac{2e_{(t)}}{rc^2}\right)\left(dx^2 + dy^2 + 
dz^2\right) , \eqno(35) $$
where $ e_{(t)} $ is the variation of the electric charge in the time.

In the Vaidya metrics [18] for scalar curvature $ R $ is valid relation [20]:

$$ R = \frac{6\dot{e}_{(t)}}{r^2c^3} = \frac{6e}{r^2ac^2}. \eqno(36) $$

\newpage
Using relations (36) the electromagnetic energy density $ \varepsilon _e $  
(34) can be expressed by the relation [20]:

$$ \varepsilon _e = -\frac{mc^4}{8\pi e}R = - \frac{3mc^2}{4\pi r^2a}. 
\eqno(37) $$

Analogically as we have determined the effective gravitation range $ r_{ef} $ 
(22) we can determine the effective range of the electromagnetic interaction 
$ r_{ef(e)} $. There holds the relation [20]:

$$ r_{ef(e)} = \sqrt{r_ea} .\eqno(38) $$

Than the Coulomb force [20]

$$ F_e = -\frac{mc^4}{4\pi e}R\frac 13\pi r_ea = -\frac{e^2}{r^2}, \eqno(39)$$
and the Coulomb potential energy [20]

$$ E_e = -\frac{e^2}{r} .\eqno(40) $$

\section{CONCLUSIONS}

Theoretically, the gravitation interaction and the electromagnetic 
interaction act in the range of the whole universe. However, in the ENU 
the gravitation interaction and the electromagnetic interaction can manifest 
itself only in limited distances given by equations (22) and (38). This is 
in good agreement with observations [17, 20].

From relations (9), (20), (24), (25), (37), (39) and (40) results the 
essential connection of the permanent constant maximum possible creation 
(actualization) of energy $ \delta c^2 $ with the gravitation interaction 
and the electromagnetic interaction in the ENU [20].

\section*{REFERENCES}

\begin{description}

\item{[1]} {\sc Einstein, A. (1915).} {\em Sitzber. Preuss. Akad. Wiss.} 
{\bf48}, 844.

\item{[2]} {\sc Schwarzschild, K. (1916).} {\em Sitzber. Preuss. Akad. Wiss.} 
 424.

\item{[3]} {\sc Friedmann, A. A. (1922).} {\em Z. Phys.} {\bf 10}, 377.

\item{[4]} {\sc Friedmann, A. A. (1924).} {\em Z. Phys.} {\bf 21}, 326.

\item{[5]} {\sc Robertson, H. P. (1929).} {\em Proc. U.S. Nat. Acad. Sci.} 
{\bf 15}, 168.

\item{[6]} {\sc Robertson, H. P. (1935).} {\em Astrophys. J.} {\bf 82}, 284.

\item{[7]} {\sc Robertson, H. P. (1936).} {\em Astrophys. J.} {\bf 83}, 187.

\item{[8]} {\sc Robertson, H. P. (1936).} {\em Astrophys. J.} {\bf 83}, 257.

\item{[9]} {\sc Walker, A. G. (1936).} {\em Proc. Lond. Math. Soc.} 
{\bf 42}, 90.

\item{[10]} {\sc Einstein, A. (1917).} {\em Sitzber. Preuss. Akad. Wiss.} 
{\bf 1}, 142.

\item{[11]} {\sc De Sitter, W. (1917).} {\em Koninklijke Academie van 
Wetenschappen te Amsterdam} {\bf 19}, 1217.

\item{[12]} {\sc Hawking, S. W. (1988). {\em A Brief History of Time: 
From the Big Bang to Black Holes,}  Bantam Books, New York.}

\item{[13]} {\sc Skalsk`y, V. (1991).} {\em Astrophys. Space Sci.} 
{\bf 176} , 313. (Corrigendum: {\sc Skalsk`y, V. (1992).} 
{\em Astrophys. Space Sci.} {\bf 187}, 163.)

\item{[14]} {\sc Skalsk`y, V. (1992). in: J. Dubni"cka (ed.),} 
{\em Philosophy, Natural Sciences and Evolution,} {\sc (Proceedings of an 
Interdisciplinary Symposium, Smolenice December 12-14, 1988), 
Slovak Academy of Science, Bratislava,} p. 83.

\item{[15]} {\sc Skalsk`y, V. (1989).} {\em Astrophys. Space Sci.} {\bf 158},
 145.

\item{[16]} {\sc Skalsk`y, V. and S`uken`ik, M. (1991).} 
{\em Astrophys. Space Sci.} {\bf 178}, 169.

\item{[17]} {\sc Skalsk`y, V. and S`uken`ik, M. (1993).} 
{\em Astrophys. Space Sci.} {\bf 209}, 131.

\item{[18]} {\sc Vaidya, P. C. (1951).} {\em Proc. Indian Acad. Sci.} 
{\bf A33}, 264.

\item{[19]} {\sc Shipov, G. I. (1993).} {\em Theory of Physical Vacuum,} 
{\sc NT-Centr, Moscow.}

\item{[20]} {\sc Skalsk`y, V. and S`uken`ik, M. (1995). {\em Proceedings of 
Scientific Works of Faculty of Sciences and Technology of the Slovak Technical 
University in Bratislava,} Trnava,} vol. {\bf 3}, p. 169.

\item{[21]} {\sc Skalsk`y, V. and S`uken`ik, M. (1991).} {\em Astrophys. 
Space Sci.} {\bf 181}, 153.

\item{[22]} {\sc Skalsk`y, V. and S`uken`ik, M. (1994).} {\em Astrophys. 
Space Sci.} {\bf 215}, 137.

\end{description}

\vspace{1cm}

\section*{FIGURE CAPTIONS}

\begin{description}

\item [Figure 1.] The gravitation interaction of two bodies in the ENU.

\end{description}


\section*{\em \Large Address:}

$^1$ Faculty of Material Sciences and Technology of the Slovak Technical 
University, Pavl`inska 16, 917 24 Trnava, Slovakia \newline
FAX: 42-805-27731, E-Mail: chamula@euba.sk

\end{document}